\documentclass[preprint2]{aastex}
\usepackage{natbib}
\usepackage{fixltx2e}
\usepackage{isotope}

\shorttitle{Cross-Field Transport of SEPs in a Fluctuating IMF}
\shortauthors{Kelly, Dalla, and Laitinen}
\begin{document}

\title{Cross-Field Transport of Solar Energetic Particles in a Large-Scale Fluctuating Magnetic Field}
\author{J. Kelly, S. Dalla, T. Laitinen }
\email{jkelly2@uclan.ac.uk}

\affil{Jeremiah Horrocks Institute, University of Central Lancashire, Preston, Lancashire, PR1 2HE, United Kingdom}

\begin{abstract}
The trajectories of Solar Energetic Particles (SEPs) in an Interplanetary Magnetic Field (IMF) exhibiting large-scale fluctuations due to footpoint motions originating in the photosphere, are simulated using a full-orbit test-particle code. The cross-field transport experienced by the particles in three propagation conditions (scatter-free, with scattering mean free path $\lambda$=0.3 AU and  $\lambda$=2 AU) is characterized in the Parker spiral geometry. The role of expansion of the magnetic field with radial distance from the Sun is taken into consideration in the calculation of particle displacements and diffusion coefficients from the output of the simulations. It is found that transport across the magnetic field is enhanced in the $\lambda$=0.3 AU and $\lambda$=2 AU cases, compared to the scatter-free case. Values of the ratios of perpendicular to parallel  diffusion coefficients vary between 0.01 and 0.08. The ratio of latitudinal to longitudinal diffusion coefficient perpendicular to the magnetic field is typically 0.2, suggesting that transport in latitude may be less efficient.  

\end{abstract}
\keywords{Sun:particle emission, Sun:heliosphere, Solar-terrestrial relations, Solar wind, Sun:activity  }

\section{Introduction}

As Solar Energetic Particles are transported through the Interplanetary Magnetic Field out to locations at 1 AU and beyond, they experience magnetic field fluctuations which occur over a range of scales. The main consequences of a particle's interaction with such fluctuations are thought to be: (a) diffusion in the direction parallel to the magnetic field, characterized by a diffusion coefficient $\kappa_{||}$, that typically describes pitch-angle scattering counteracting the strong magnetic focusing in the IMF and (b) diffusion in the direction perpendicular to the average field, characterized by a coefficient $\kappa_{\perp}$, thought to be orders of magnitude smaller than $\kappa_{||}$.

Diffusion in the parallel direction has been the subject of a large number of studies over the past three decades, relevant not only to SEPs but also cosmic rays and more general particle propagation in turbulent fields. While the initial consensus was that diffusion was the main factor determining profiles of SEP intensity versus time, a paradigm shift took place, that implied the role of interplanetary transport in shaping SEP profiles was minimal \citep{1999SSRv...90..413R}. More recent studies, however, have re-affirmed that diffusion does play a role in determining SEP profiles \citep{2006ApJ...647L..65M}.

For SEPs, diffusion in the perpendicular direction has been largely ignored until recently, and is currently the subject of much interest and controversy. Some of the observational studies pointing towards perpendicular transport playing an important role, used data at high heliolatitudes from Ulysses \citep{2009ApJ...692..109Z, 2003AnGeo..21.1367D, 2003GeoRL..30sULY9D}. Recent observations from STEREO show that electrons in \isotope[3]{He}-rich events reach spacecraft widely separated in longitude, a fact that may require transport across the field \citep{2010AIPC.1216..621W}. A number of theoretical studies have also addressed perpendicular diffusion for SEPs, as detailed below.

A common technique when modelling energetic particle propagation is to adopt a kinetic approach with a diffusion tensor including parallel and perpendicular diffusion coefficients, $\kappa_{||}$ and $\kappa_{\perp}$. In this approach $\kappa_{||}$ and $\kappa_{\perp}$ are an input to the model and are estimated using observational data. Particle gyromotion is neglected and a kinetic transport equation is solved numerically to give a gyro-averaged phase-space density function for the SEP population. 

Such an approach is taken by \citet{2009ApJ...692..109Z} where the Fokker-Planck equation is recast into a set of stochastic differential equations (SDEs) that are solved numerically to investigate the propagation of high-energy SEPs. They found that the inclusion of perpendicular diffusion increased the uniformity of particle flux when calculated at different locations during the decay phase of a simulated event. This phenomenon has been observed in SEP events and is refered to as the ``reservoir phenomenon'' \citep{1972JGR....77.3957M}.  

A similar approach is taken in \citet{2011ApJ...734...74H} where the influence of SEP source characteristics on observations at 1AU is investigated. They found that perpendicular diffusion plays a significant role in SEP propagation, particularly in cases where a spacecraft is not directly connected to the acceleration region; in such cases, the earliest arriving particles can be seen propagating towards the Sun, having scattered backwards at large distances. 

\citet{2010ApJ...709..912D} solved the Fokker-Planck equation using a time-forward SDE method with two different particle scattering schemes, to investigate the spatial distribution of SEPs in the inner heliosphere. They found that scaling the value of $\lambda_{\perp}$ - the mean free path in the direction perpendicular to the field - with gyroradius gave more realistic spatial distributions at $1$ AU than when $\lambda_{\perp}/\lambda_{||}$ was kept constant at a value of $0.01$. 

Another technique that can be used to understand the cross-field transport of SEPs is a full-orbit test-particle simulation. In this scheme a particle's equations of motion are solved numerically to determine its trajectory. Such an approach assumes that test-particles do not interact with one another, or affect the IMF.

\citet{2011JGRA..11602102T} used a full-orbit test-particle method to investigate the properties of SEP transport in a model of the IMF that included isotropic small-scale turbulence. They found that such turbulence led to diffusive behaviour and that $\lambda_{\perp}/\lambda_{||}=0.1$ at around 10AU, slightly larger than observed values.

In this paper, the propagation of SEPs in large-scale turbulence is investigated by means of full-orbit test-particle simulations. The turbulence is induced by footpoint motion associated with supergranulation, leading to large-scale fluctuations, according to the description first developed by \citet{1999AdSpR..23..581G} to investigate the propagation of CIR-associated particles to high heliolatitudes. He ran full-orbit test-particle simulations of $1$ MeV protons, injected at $2$ AU and integrated for a period of $10$ days, and found that field line braiding resulting from magnetic footpoint motion provides an efficient mechanism for latitudinal transport. 

\citet{2006ApJ...641.1222P} used a test-particle method to investigate the onset times of SEPs in the IMF model put forward by \citet{1999AdSpR..23..581G}. They found that such fluctuations can reduce the length of magnetic field lines and hence lead to onset times for SEP events shorter than within a Parker spiral model.

Here the effects of large-scale turbulence on the spatial distribution of SEPs, in particular its role spreading particles along the direction perpendicular to the magnetic field, are investigated. Three propagation regimes are analysed: a scatter-free regime and two scattering ones with different mean free paths. Diffusion coefficients are calculated from the output of the test particle model and the influence of the expansion of Parker spiral magnetic field lines on their values and evolution discussed.

An overview of the IMF model and the parameters used in simulations is presented in Section \ref{sec.model}, along with details relating to the test-particle method's implementation. In Section \ref{sec.results} the results of scatter-free simulations are presented and contrasted with corresponding simulations which include scattering. Finally in Section \ref{sec.discuss} the results are summarized and conclusions are outlined.

\section{Model}
\label{sec.model}

\subsection{Interplanetary Magnetic Field}
\label{subsec.IMF}
In this work the IMF model originally introduced by \citet{1999AdSpR..23..581G}, and further developed by \citet{2004ApJ...616..573G}, is used. Large-scale magnetic field fluctuations are generated by supergranular motion of the footpoints at the solar surface, with characteristic time $T_c$ and speed $V_g$.
 
The resulting fluctuating IMF is given by:
\begin{equation}B_{r}=\frac{B_{0}r_{0}^{2}}{r^{2}}  \label{eq.br} \end{equation}

\begin{equation}B_{\theta}=\frac{B_{0}r_{0}^{2}}{r^{2}}\frac{V_{\theta}(\theta,\phi,t_{0})}{V_{w}}\end{equation}

\begin{equation}B_{\phi}=\frac{B_{0}r_{0}^{2}}{r^{2}}\frac{V_{\phi}(\theta,\phi,t_{0}) - r\Omega_{0}\sin{\theta}}{V_{w}}  \label{eq.bphi} \end{equation}

\begin{equation} t_{0}=t - \frac{r-r_{0}}{V_{w}}\end{equation}
where $V_{\theta}(\theta,\phi,t_{0})$ and $V_{\phi}(\theta,\phi,t_{0})$ are velocities associated with footpoint motion, $V_{w}$ is the solar wind speed, $\Omega_{0}$ is the solar rotation rate, $B_{0}$ is the magnetic field magnitude at position $r_{0}$, $r$ is radial distance from the Sun, $\theta$ is colatitude, $\phi$ is longitude, $t$ is time, and $t_{0}$ represents the time at which a fluctuation at a given radius was produced on the solar surface. 

The velocity of the footpoint motions is modelled by a stream function $\psi(\theta, \phi, t_{0})$, satisfying:

\begin{equation}
V_{\theta} = \frac{1}{\sin \theta} \frac{\partial \psi}{\partial \phi}
\label{eq.vtheta}
\end{equation}

\begin{equation}
V_{\phi}= - \frac{ \partial \psi }{\partial \theta}
\label{eq.vphi}
\end{equation}

The stream function is postulated to be:
\begin{equation}
\psi(\theta, \phi, t_{0}) = \sum_{n=1}^{N} \sum_{m=-n}^{n} a_{n}^{m} \exp(i\omega_{n}^{m} t_{0}+ i \beta_{n}^{m}) Y_{n}^{m}(\theta, \phi)
\label{eq.stream}
\end{equation}
where:
\begin{equation}
a_{n}^{m} = 6 V_{g}G_{n}^{m}\left( \sum_{n=1}^{N} \sum_{-n}^{n} G_{n}^{m} \right)^{-1}
\label{eq.amp}  
\end{equation}

\begin{equation}
G_{n}^{m} = \left(\left[1 + \left(\frac{n}{N_{c}}\right)^{10/3} \right] \left[1+ (\omega_{n}^{m} T_{c})^{5/3} \right]\right)^{(-1/2)}
\label{eq.g}
\end{equation}
$N_{c}$ is the characteristic number of supergranular cells, defined as $N_{c}=\pi r_{0}/V_{g} T_{c}$, where $r_{0}$ is taken to be one solar radius, r\textsubscript{s}. $\omega_{n}^{m}$ is the characteristic frequency of each mode and is taken as $\omega_{n}^{m} = 2 \pi n/150 T_{c}$. The remaining parameters in the stream function are: random phase angles, $\beta_{n}^{m}$, drawn from a uniform distribution, and $ Y_{n}^{m}(\theta, \phi) $ the spherical harmonic function. The values of various parameters used in these simulations are presented in Table \ref{tab.param}.

\begin{table}
\begin{center}
\begin{tabular}{ c|c }

Parameter & Value \\
\hline
\hline
$B_{0}$ (G) & $1.78$ \\
$V_{w}$ (cm s\textsuperscript{-1}) & $5.0\times10^{7}$  \\
$\Omega_{0}$ (rad s\textsuperscript{-1}) & $2.86533\times10^{-6}$  \\
$T_{c}$ (day) & $1$  \\
$V_{g}$ (km s\textsuperscript{-1})& $2.0$  \\
$N$ & $50$ \\
$r_0$ (r\textsubscript{s}) & 1  \\

\end{tabular}
\end{center}
\caption{Values of the parameters used in all the simulations. The values of $T_{c}$, $v_{g}$ and $N$ are those used in \citet{2006ApJ...641.1222P}.}
\label{tab.param}
\end{table}

The electric field is given by $\mathbf{E}=-\mathbf{V}_w \times \mathbf{B}/c$ since the plasma is quasineutral.

In Section \ref{sec.results} the cartesian system used is defined as follows: the $z$ axis is the rotational axis of the Sun, while the $x$ and $y$ coordinates lie in the heliospheric equatorial plane with the $x$ axis corresponding to a longitude of $\phi=0^{\circ}$.

In the simulations presented here all particles are run through a realization of this IMF model. We set $t=0$, giving $t_{0}=-(r-r_{0})/V_{w}$, as the timescale of variation of the large scale turbulence is large compared to the particle propagation times considered.

\subsection{Particle Simulation}
\label{subsec.part_sim}

The simulations presented in this paper use a full-orbit test-particle method, which involves numerically integrating the equations of motion of one particle at a time, repeating the process for a variety of initial conditions to assess the propagation of a population. 

The code is fully relativistic and is a modified version of one previously used to study particle acceleration during magnetic reconnection \citep{2005A&A...436.1103D}.
The numerical technique used for integration is a Bulirsch-Stoer method \citep{Press:1993:NRF:563041}. The Bulirsch-Stoer routine is driven by a function which adaptively controls the stepsize of integration, so as to produce results to a prescribed accuracy with minimum computational effort. The solution's accuracy is determined by a user-specified tolerance value, which sets the fractional error allowed when integrating from $t_{n}$ to $t_{n+1}$. 

Extensive testing was carried out to find an acceptable tolerance level for the solver; from these tests it was concluded that particle trajectories in the fluctuating IMF outlined in Section \ref{subsec.IMF} converged when the tolerance was less than $10^{-11}$, hence the simulations were carried out at a tolerance of $10^{-12}$.  

The presence of small-scale turbulence in the solar wind affects the trajectories of particles. Rather than providing a full description of the interaction of particles with turbulence in these simulations, its effect is described as a series of scattering events, each causing an instantaneous change in a particle's velocity. This `ad-hoc' scattering has been used by many others in the literature (eg \citet{2006ApJ...641.1222P}).

A mean free path, $\lambda$, is fixed and the mean scattering time $t_{scat}$ is obtained by $t_{scat}= \lambda / v$, where $v$ is the particle's velocity in a frame of reference which is stationary with respect to the solar wind. A series of scattering events are then numerically generated for each particle, with mean time $t_{scat}$, in such a way as to form a Poisson distribution.
A scattering event involves firstly transforming the particle's velocity vector into the solar wind frame, then the particle's velocity vector is reassigned by drawing a velocity vector at random from an isotropic distribution, changing both the particle's pitch- and phase-angles. The particle's new velocity is then converted back into the inertial frame of reference, and the integration of the particle's equations of motion continues. 

\section{Results}
\label{sec.results}

In all simulations presented in this paper a mono-energetic population of protons, with velocities randomly distributed in a hemisphere directed away from the Sun, is injected instantaneously into the interplanetary medium. The location of injection is a $3^{\circ} \times 3^{\circ}$ area on a spherical surface of radius $21.5$ $r_{s}$ centred on a colatitude of $60^{\circ}$ and a longitude of $1.5^{\circ}$.

\subsection{Scatter-free Propagation}
\label{subsec.scat-free}

Initial simulations were performed without ad-hoc scattering in order to investigate the cross-field transport of SEPs subjected only to large-scale fluctuations, and contrast their propagation in such a field with that in a Parker spiral field.

\begin{figure*}
\epsscale{1.2}
\plotone{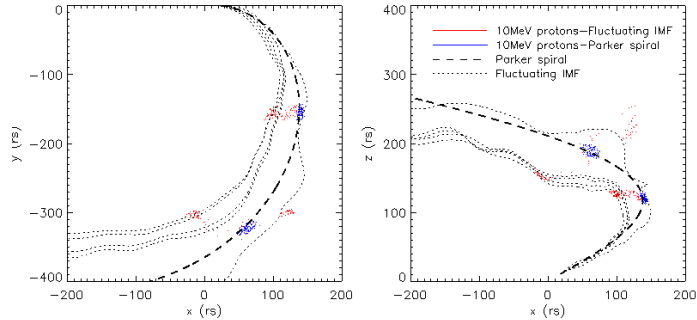}
\caption{An $x$-$y$ (left) and $x$-$z$ (right) projection of 10 MeV proton locations at $t$=30 mins  and $t$=60 mins after injection for scatter-free propagation in fluctuating IMF (red dots) and Parker IMF (blue dots). The thick dashed line is a Parker spiral line from the centre of the injection region, and dotted lines are field lines of the fluctuating IMF with starting locations at the center and corners of the injection region.  \label{fig.noscat_xy_xz}} 
\end{figure*}

Figure \ref{fig.noscat_xy_xz} shows the spatial distributions of $10$ MeV protons $30$ and $60$ minutes after injection. The left panel shows an $x$-$y$ projection (along the heliospheric equatorial plane) and the right panel an $x$-$z$ projection. Red dots are particles in the fluctuating IMF configuration and blue dots those in the Parker spiral (long dashed line). Field lines associated with the fluctuating IMF are indicated by short dashed lines. It can be seen that the longitudinal and latitudinal spread of SEPs is much greater in the fluctuating IMF model than in the Parker spiral.

\begin{figure}
\epsscale{0.8}
\plotone{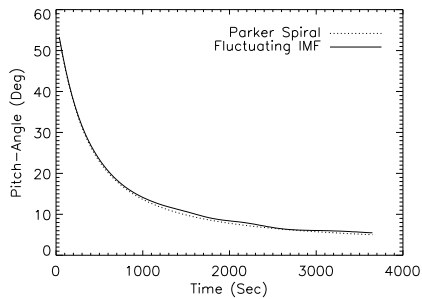}
\caption{A plot of average pitch-angle vs time for all particles in the Parker spiral (dashed line) and the fluctuating IMF (solid line) simulations. The proton energy is 10 MeV and the distance from the Sun at the final time is about 215 $r_s$. \label{fig.pitch_comp}} 
\end{figure}

The smallest scale of the fluctations is $2\pi r_{s} / 50 = 87$ Mm, this is large compared to the gyroradii of SEPs propagating close to $1$ AU, hence the cross-field transport observed in the simulation results is simply due to the field line meandering and drifts. The particles propagating in the fluctuating IMF model focus at the same rate as those in the Parker spiral, as can be seen in Figure \ref{fig.pitch_comp}, showing the evolution of the mean pitch angle of the populations with time. 

To investigate any dependence of SEP propagation characteristics in the fluctuating IMF on particle energy, proton energies were varied from $500$ KeV to $1$ GeV. Results showed that SEP populations followed the same path through a single realization of the fluctuating IMF regardless of energy, and hence cross-field transport in this configuration is due solely to field line meandering.

\subsection{Propagation with Scattering}
\label{subsec.scat}

The scatter-free simulation described in Section \ref{subsec.scat-free} showed that in such conditions cross-field transport in the fluctuating IMF is due solely to field-line meandering. In this Section the effect of ad-hoc pitch-angle scattering, incorporated into the model as outlined in Section \ref{subsec.part_sim}, is studied.

\begin{figure*}
\epsscale{2.0}
\plotone{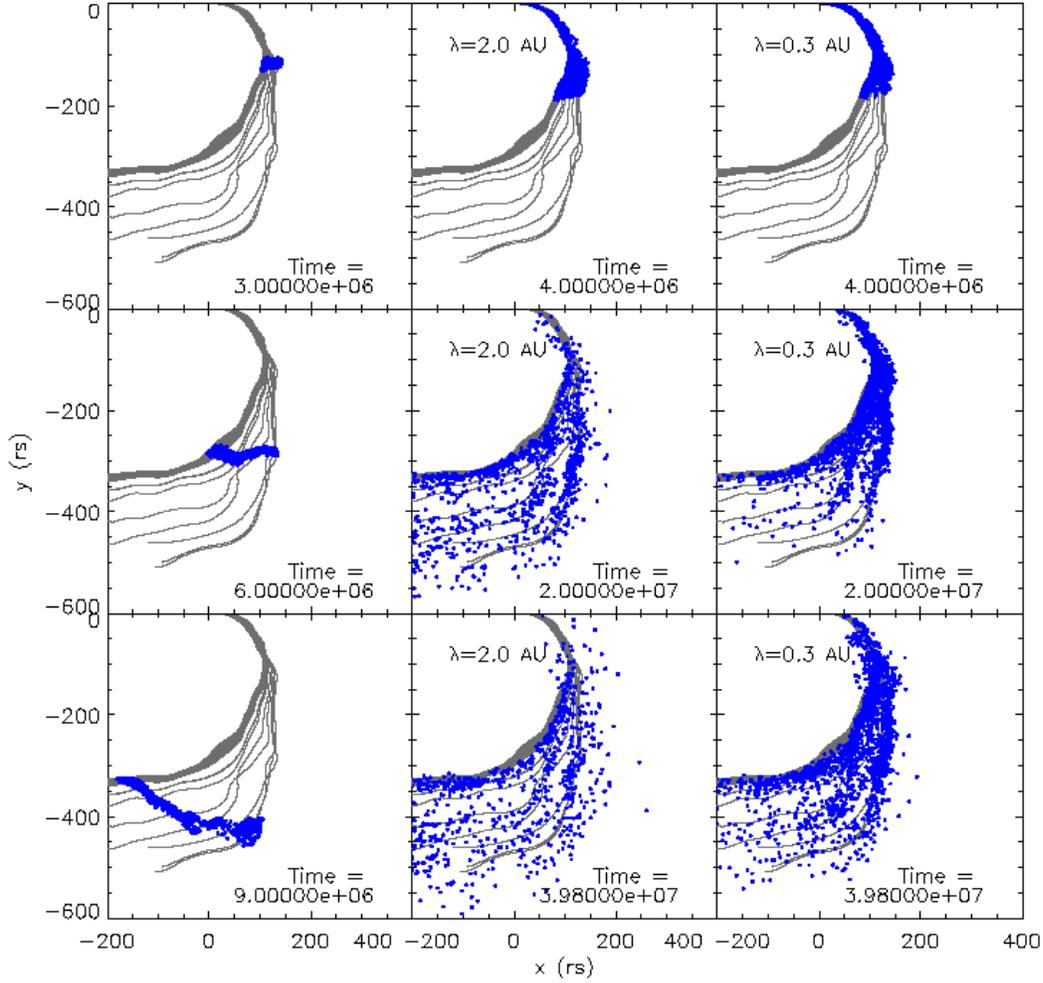}
\caption{$x$-$y$ projections of a population of 2000 $100$ MeV protons propagating in the fluctuating IMF at three different times. The left column is for scatter-free propagation, the central one for ad-hoc pitch-angle scattering with $\lambda = 2.0$ AU, and the right one for scattering with $\lambda = 0.3$ AU. Times indicated are dimensionless, but can be converted to hours using $1.0e+7 = 1$ hour. The grey lines are magnetic field lines with seed points on a grid overlaid on the injection region. For the $\lambda = 2.0$ AU case, many particles have propagated to large radial distances and are not shown in the plot.  \label{fig.xy_xz_stack}} 
\end{figure*}

Figure \ref{fig.xy_xz_stack} shows populations of $100$-MeV protons propagating in the fluctuating IMF. Each column shows a particle population at three different times for a given propagation condition. The initial conditions are the same in the three cases. The central and right columns are affected by pitch angle scattering, with mean free paths of $2.0$ AU and $0.3$ AU respectively, and the left column shows scatter-free propagation. In the scattered populations there are a number of particles close to their injection point at late times, in addition to a number of outlying particles that have strayed a large distance across the field compared to the scatter-free case. 

To quantify the cross-field transport within the simulations, the parallel and perpendicular displacements of the particles' location from the Parker spiral field line originating from their initial positions were calculated, for runs of 10000 protons of 50 MeV energy.
The first step consists of deriving the so-called \lq target point\rq, i.e. the location on the Parker spiral field line starting at the particle's initial position, located nearest to the actual final position \citep{2011JGRA..11602102T} (see Appendix \ref{app.targetradius} and Fig.~\ref{fig.target}). If $\mathbf{r}$ indicates the location of a particle at a given time and $\mathbf{r}_t$
is the location of the target point, the displacement of the particle with respect to its target location is $\Delta \mathbf{s} = \mathbf{r}-\mathbf{r}_t$. Next a local Parker spiral coordinate system, centered in the target location, is introduced, with an axis
pointing outwards along the spiral (the \lq parallel\rq\ direction), a second axis in the direction of $\mathbf{e}_{\theta'}$=$-$$\mathbf{e}_{\theta}$ with $\mathbf{e}_{\theta}$ the standard spherical coordinate system unit vector and a third axis $\mathbf{e}_{\phi'}$  completing the orthogonal system (see Appendix \ref{app.transform}). The components of $\Delta \mathbf{s}_{\perp}$ in this system are $\Delta s_{\phi^{\prime}}$ and $\Delta s_{\theta^{\prime}}$. The distance travelled along the Parker spiral between the initial location and the target point is indicated as $l_{||}$.  

The distributions of displacements for the three propagation conditions of Figure \ref{fig.xy_xz_stack} are examined first after a fixed time from injection (when the average radial distance from the Sun in the population will be different in the three cases) and second at the times when the average distance from the Sun is the same for the three populations  and set to 1 AU.

\begin{figure}
\epsscale{0.8}
\plotone{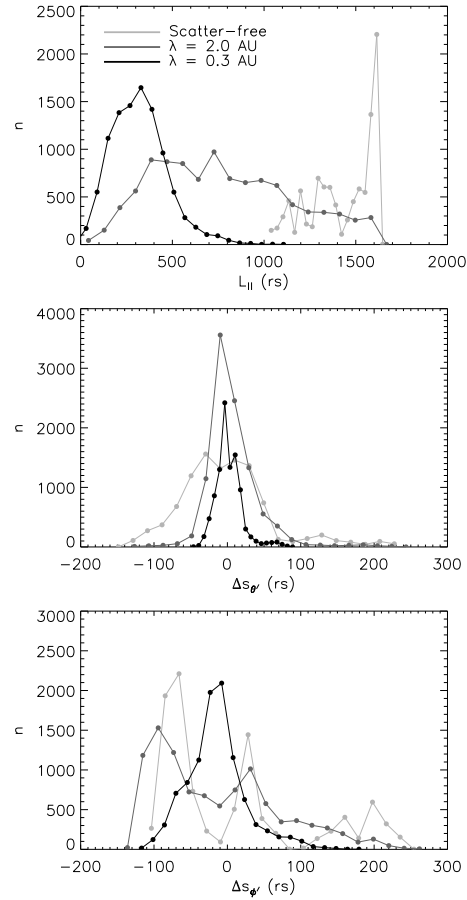}
\caption{$l_{||}$ (top), $\Delta s_{\theta^{\prime}}$ (middle) and $\Delta s_{\phi^{\prime}}$ (bottom) distributions of $50$ MeV protons at a time corresponding to around 3 hours. Unscattered population: light gray, $\lambda=2.0$ AU population: gray, and $\lambda=0.3$ AU population: black. \label{fig.ds_directions_t}} 
\end{figure}

Figure \ref{fig.ds_directions_t} shows the distributions of the displacement's components for the scatter-free, $\lambda=2$ AU and $\lambda=0.3$ AU conditions, for $50$ MeV protons, three hours after injection. The top panel, displaying the distribution of $l_{||}$, shows that the population with $\lambda=0.3$ AU has traveled the shortest distance along the field, followed by the $\lambda=2.0$ AU case, and that the unscattered case has traveled the furthest, as one would expect. 
This difference in average field-parallel displacement affects the magnitude of perpendicular displacement, as the spread of the field lines in the perpendicular direction increases with distance along the field. This leads to the distributions in the middle and bottom panels of Figure \ref{fig.ds_directions_t}, where the scatter-free population has wide distributions since all these particles have been able to travel a greater distance along the field.
 


In an attempt to remove the effect of field line expansion on $\Delta s_{\perp}$ plots, $\Delta s_{\theta^{\prime}}$ and $\Delta s_{\phi^{\prime}}$ values were normalised by the scale lengths of a Parker spiral flux-tube at the corresponding distance from the Sun, as defined in Appendix \ref{app.fluxnorm}. Figure \ref{fig.ds_directions_norm_t} shows the normalized distributions. In the $\theta^{\prime}$ component, the unscattered population still has the widest distribution, while in the $\phi^{\prime}$ component the widest distribution is in the $\lambda$=2 AU case.    
It should be noted that the normalisation introduced only corrects for the expansion of the Parker spiral field lines: the superimposed large scale turbulence may be characterized by a different kind of expansion which is not accounted for here. 

\begin{figure}
\epsscale{0.8}
\plotone{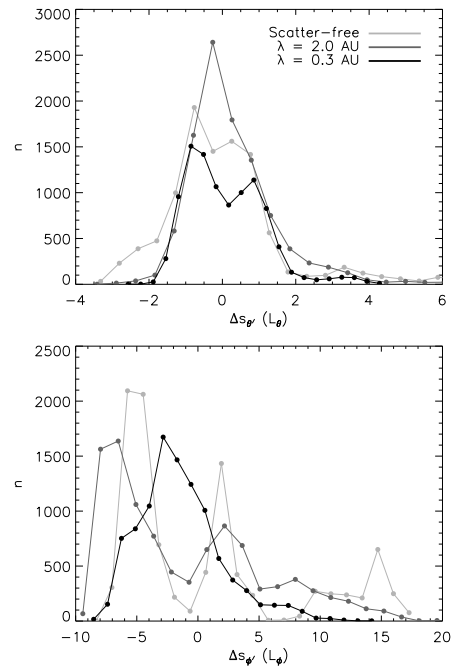}
\caption{ $\Delta s_{\theta^{\prime}}$ (top) and $\Delta s_{\phi^{\prime}}$ (bottom) distributions of $50$ MeV protons, normalized by Parker spiral flux tube widths $l_{\theta}$ and $l_{\phi}$ respectively, as defined in Section \ref{app.fluxnorm}, at a time corresponding to around 3 hours. Propagation conditions are indicated as in Figure \ref{fig.ds_directions_t}.  \label{fig.ds_directions_norm_t}} 
\end{figure}

\begin{figure}
\epsscale{0.8}
\plotone{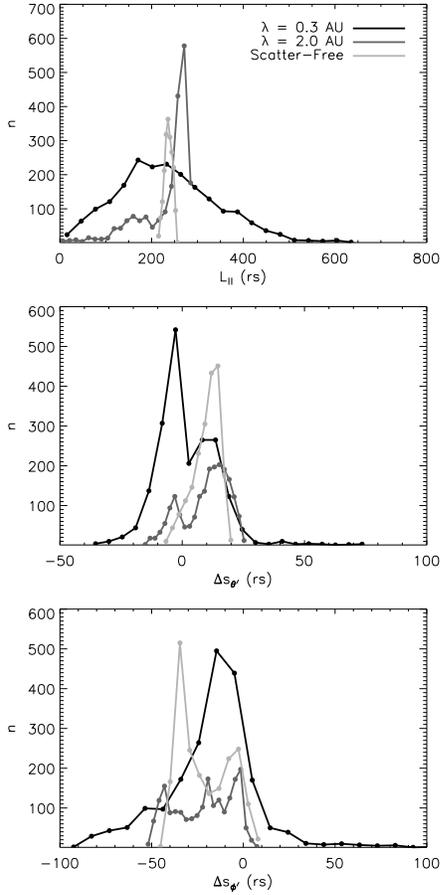}
\caption{ $l_{||}$ (top), $\Delta s_{\theta^{\prime}}$ (middle) and $\Delta s_{\phi^{\prime}}$ (bottom) distributions of $50$ MeV protons at the time when the average radius of each population is $1$ AU. Propagation conditions are indicated as in Figure \ref{fig.ds_directions_t}.  \label{fig.dir_r_dist} } 
\end{figure}

Figure \ref{fig.dir_r_dist} compares $\Delta \mathbf{s}$ distributions at the time when each population has an average radial distance from the Sun of $1$ AU. For the scatter-free case, $\lambda=2$ AU and the $\lambda=0.3$ AU cases this is 29, 33 and 91 minutes after injection respectively. Particles are grouped in this way in an attempt to find the most appropriate time to compare populations with different scattering conditions. Populations with little scattering will reach greater radial distances in a given time than populations with strong scattering, and at larger radial distances the field line expansion due to magnetic field fluctuations will be larger, artificially increasing the the $\Delta s_{\perp}$ value assigned to a particle  

Grouping particles this way results in the $\lambda=0.3$ AU population having the broadest distribution in both perpendicular directions, however that population has travelled the furthest along the field (see $l_{||}$ distribution in the top panel of Figure \ref{fig.dir_r_dist}) and this produces the pronounced \lq wings\rq\ in the middle and bottom panel of Figure \ref{fig.dir_r_dist}. This effect is similar to that seen in Figure \ref{fig.ds_directions_t} and is due to field line expansion.



\begin{figure}
\epsscale{0.8}
\plotone{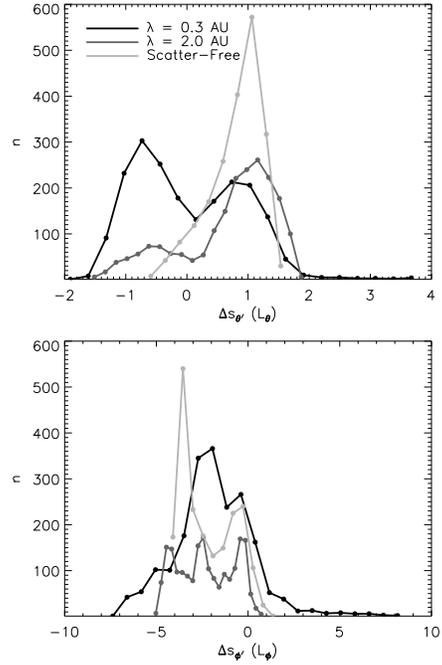}
\caption{ $\Delta s_{\theta^{\prime}}$ (top) and $\Delta s_{\phi^{\prime}}$ (bottom) distributions for $50$ MeV protons at the time when the average radius of each population is $1$ AU, normalized by the Parker spiral flux tube widths defined in Section \ref{app.fluxnorm}. Propagation conditions are indicated as in Figure \ref{fig.ds_directions_t}. \label{fig.dir_r_dist_norm} } 
\end{figure}

Figure \ref{fig.dir_r_dist_norm} shows the $\Delta s_{\perp}$ components normalized by the Parker spiral flux-tube scale lengths. The $\lambda=0.3$ AU population has the broadest distribution in both the $\phi^{\prime}$ and $\theta^{\prime}$ populations. The distributions for the scattered populations are broader than for the scatter-free one. 

\begin{figure}
\epsscale{0.8}
\plotone{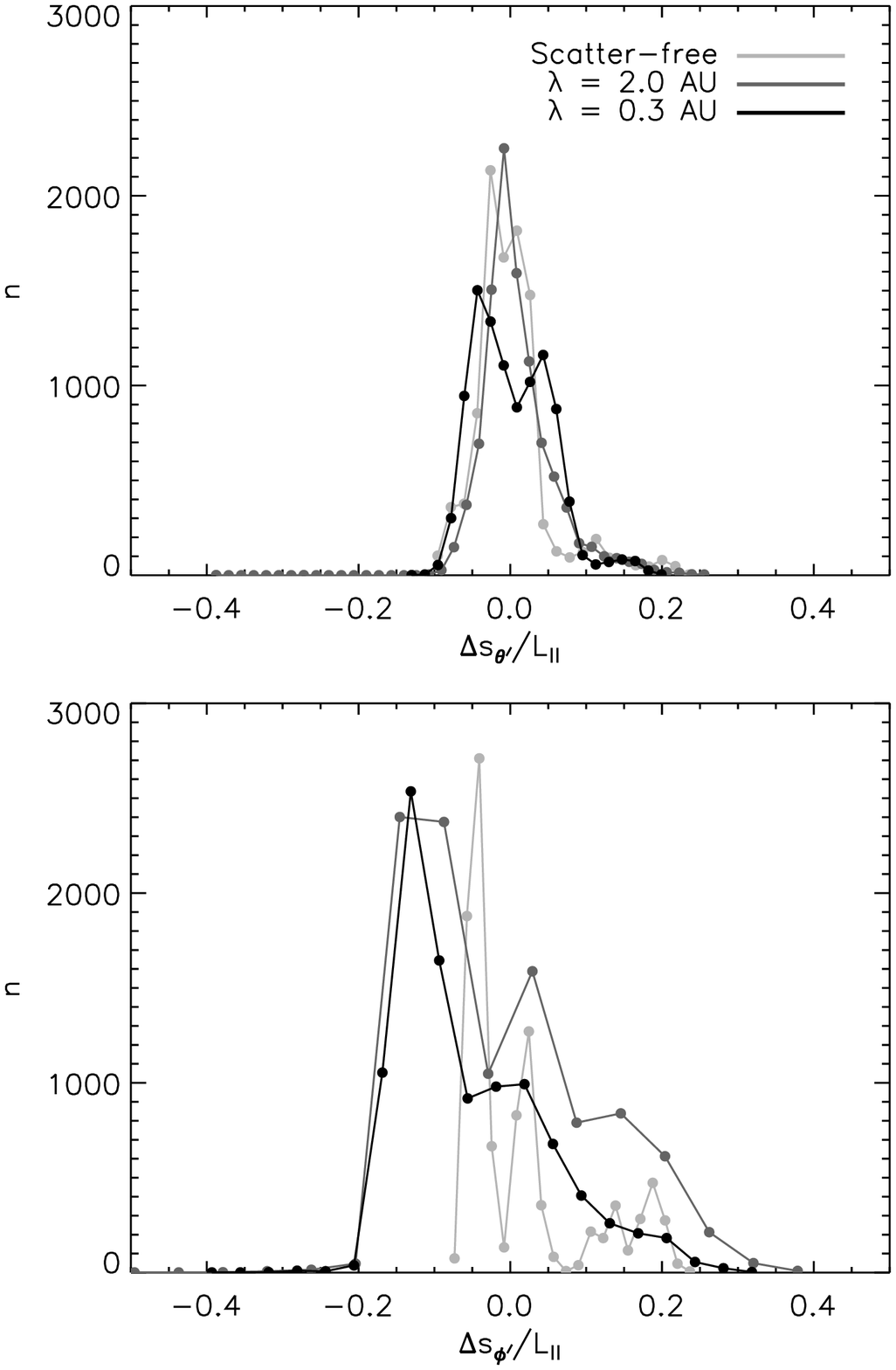}
\caption{ $\Delta s_{\theta^{\prime}} / l_{||} $ (top) and $\Delta s_{\phi^{\prime}} / l_{||} $ (bottom) distributions of $50$ MeV protons at a time corresponding to around 3 hours. Propagation conditions are indicated as in Figure \ref{fig.ds_directions_t}. \label{fig.ds_per_o_par_t}} 
\end{figure}

An alternative way to remove the effect of field line expansion on $\Delta s_{\perp}$ values involves normalizing $\Delta s_{\perp}$ by $l_{||}$ for each particle in the population at a fixed time 3 hours after injection, giving the cross-field displacement per unit distance along the field. The results of this normalization are presented in Figure \ref{fig.ds_per_o_par_t}, which shows that the population with $\lambda=2.0$ AU has the broadest distribution of $\Delta s_{\phi^{\prime}}/l_{||}$ and has many outliers in the plot of $\Delta s_{\theta^{\prime}}/l_{||}$. This result ties in with Figure \ref{fig.xy_xz_stack}, where the $\lambda=2.0$ AU population qualitatively appears to have outliers that have strayed the largest distance across the field.


Particle transport can be characterized by diffusion coefficients, under the assumption that it is of diffusive character. 
The values of diffusion coefficients  $\kappa_{||}$, $\kappa_{\phi^{\prime}}$  and $\kappa_{\theta^{\prime}}$ are calculated from the equation:
\begin{equation}
\kappa_{||, \phi^{\prime}, \theta^{\prime} } = \frac{ \langle(\Delta s_{||, \phi^{\prime}, \theta^{\prime}})^{2}\rangle}{2t}
\label{eq.diff_coeffs}
\end{equation}
In the expanding magnetic field the propagation parallel to the field is a combination of diffusion and focusing-driven
streaming. This can be represented as advection of a diffusively-spreading particle population, in the strong scattering case
\citep{Earl1976}. In line with this concept, the parallel displacement  $\Delta s_{||}$ is defined relative to the mean position of the particle distribution as: 
\begin{equation}
\Delta s_{||}=l_{||}-\langle l_{||} \rangle
\end{equation}
where the average is taken over the particle population.

This definition of diffusion coefficient includes the contribution of displacements in the particle's trajectory due to large-scale fluctuations in the magnetic field, and as such may not be suitable to directly compare with the diffusion coefficient used in the Parker transport equation.

\begin{figure}
\epsscale{0.8}
\plotone{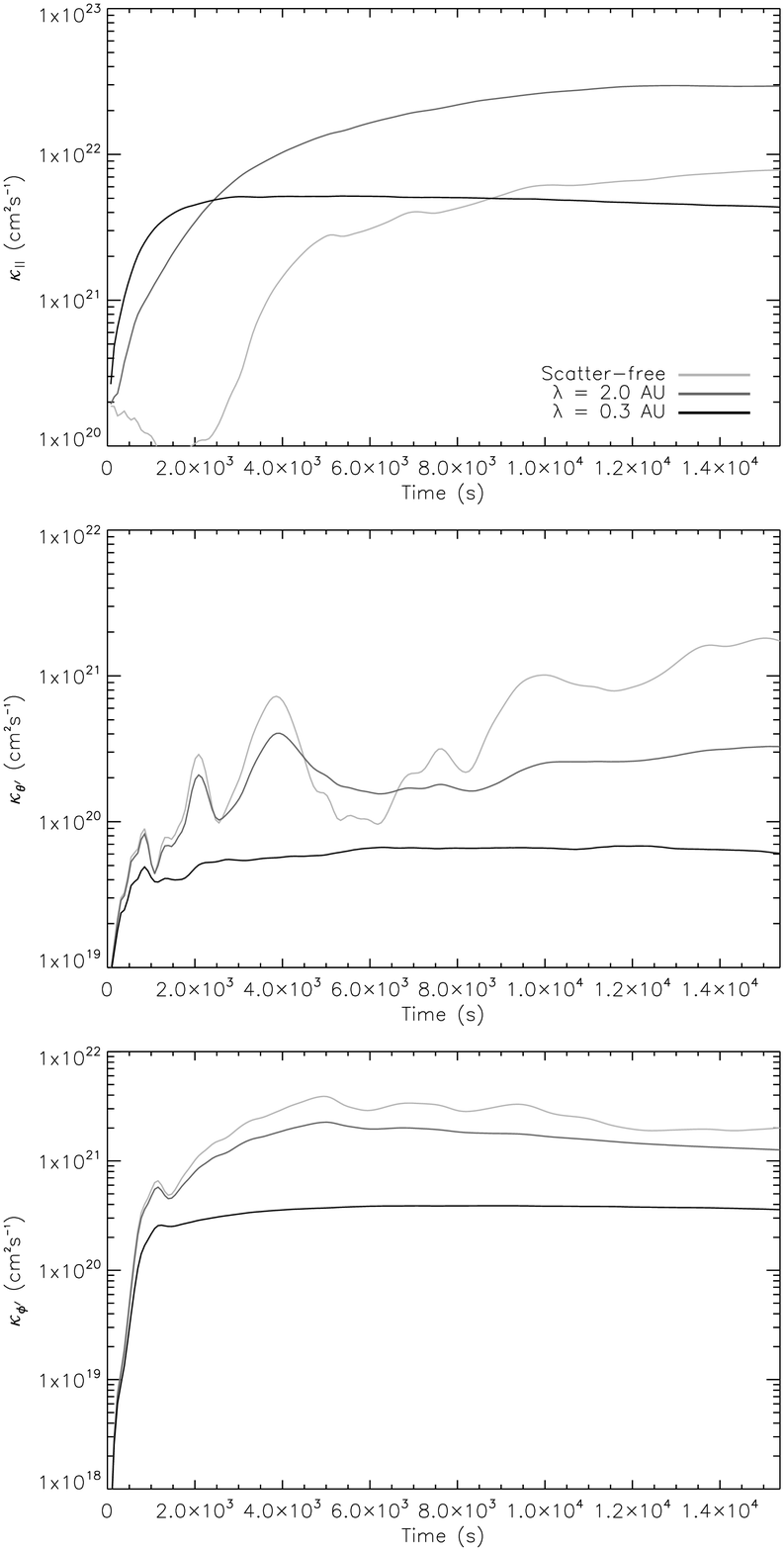}
\caption{$\kappa_{||}$ (top), $\kappa_{\theta^{\prime}}$ (middle), and $\kappa_{\phi^{\prime}}$ (bottom) running diffusion coefficients for $50$ MeV protons under different propagation conditions. Propagation conditions are indicated as in Figure \ref{fig.ds_directions_t}.  \label{fig.diff_coeffs}} 
\end{figure}

Figure \ref{fig.diff_coeffs} shows the diffusion coefficients calculated from Eq.~\ref{eq.diff_coeffs} for the three propagation conditions as a function of time. For diffusive behaviour it is expected that the diffusion coefficient would reach a constant value. The top panel shows  $\kappa_{||}$ for the three populations, although the $\kappa_{||}$ values of the unscattered population are included only for reference, as this case is not diffusive since the particles are rapidly focussed, and travel as a beam. The increase in $\kappa_{||}$ with time for the scatter-free case is due to field line meandering.

The slopes of plots shown in Figure \ref{fig.diff_coeffs} can be characterized by the relative differences between values at the midpoint and final time point in the simulations, as given in Table \ref{tab.delta_diff}. In the $\lambda=2.0$ AU case $\Delta \kappa_{||}/\kappa_{||}$=0.28, whereas for $\lambda=0.3$ AU  $\Delta \kappa_{||}/\kappa_{||}$=$-$0.16, so neither curve is constant. For the perpendicular components of the diffusion coefficient an approximately constant value is reached only in the   $\lambda=0.3$ AU case.

\begin{table}
\begin{center}
\begin{tabular}{ c|c|c }

 & $\lambda=2.0$ AU & $\lambda=0.3$ AU \\
\hline
$\Delta\kappa_{||}/\kappa_{||}$  & $0.284$ & $-0.163$ \\
$\Delta\kappa_{\phi^{\prime}}/\kappa_{\phi^{\prime}}$  & $-0.506$ & $-0.080$ \\
$\Delta\kappa_{\theta^{\prime}}/\kappa_{\theta^{\prime}}$  & $0.454$ & $-0.083$ \\

\end{tabular}
\end{center}
\caption{The difference between the midpoint and endpoint values of the diffusion coefficients, normalised by the final value, for different propagation conditions.}
\label{tab.delta_diff}
\end{table}

The ratios of various diffusion coefficients from each population are presented in Table \ref{tab.ratio_diff}. It can be seen that in both scattering conditions the value of $\kappa_{\theta^{\prime}}/\kappa_{||}$ is about 0.01, while $\kappa_{\phi^{\prime}}/\kappa_{||}$ is approximately 0.08 for $\lambda=0.3$ AU and 0.04 for $\lambda=2.0$ AU. The ratio of the two perpendicular components of the diffusion coefficient is not one, as $\kappa_{\theta^{\prime}}/\kappa_{\phi^{\prime}}$$\sim$0.2. This difference may be due to the fact that electric field drift aids diffusion in the $\mathbf{e_{\phi^{\prime}}}$ direction \citep{1968JGR....73.7377B}, or may be a property of the magnetic field meandering.  

The increased value of $\kappa_{\phi^{\prime}}$ compared with $\kappa_{\theta^{\prime}}$ found in these simulations differs from the conclusion of \citet{1995GeoRL..22.3385J}, based on analysis of magnetic field fluctuations in Ulysses data. However this difference may be explained by the fact that the Ulysses measurements were made at a much larger radial distance from the Sun. 

\begin{table}
\begin{center}
\begin{tabular}{ c|c|c }

 & $\lambda=2.0$ AU & $\lambda=0.3$ AU  \\
\hline
$\kappa_{\theta^{\prime}}/\kappa_{||}$  & $0.011$ & $0.0139$ \\
$\kappa_{\phi^{\prime}}/\kappa_{||}$  & $0.043$ & $0.083$ \\
$\kappa_{\theta^{\prime}}/\kappa_{\phi^{\prime}}$  & $0.259$ & $0.169$ \\

\end{tabular}
\end{center}
\caption{Diffusion coefficient ratios in different propagation conditions. The values of each diffusion coefficient at the final time were used to calculate these ratios.}
\label{tab.ratio_diff}
\end{table}



\begin{figure}
\epsscale{0.8}
\plotone{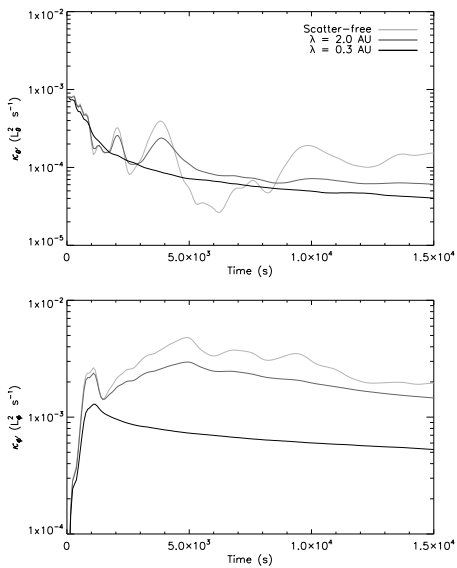}
\caption{$\kappa_{\theta^{\prime}}$ (top) and $\kappa_{\phi^{\prime}}$ (bottom) running diffusion coefficients for $50$ MeV protons, normalized by the width of a Parker spiral flux tube, outlined in Appendix \ref{app.fluxnorm}. Propagation conditions are indicated as in Figure \ref{fig.ds_directions_t}. \label{fig.diff_coeffs_direction_norm} } 
\end{figure}

The diffusion coefficients were also calculated by means of values of $\Delta s_{\theta^{\prime}}$ and  $\Delta s_{\phi^{\prime} }$ normalised by the width of a Parker spiral flux tube, as outlined in Section \ref{app.fluxnorm}; the plots of the resulting diffusion coefficients are shown in Figure \ref{fig.diff_coeffs_direction_norm}. The observed decrease in the diffusion coefficients with time can be understood by noting that in a simple radially expanding field without scattering, field line expansion would generate a dependence like $t^{-1}$ for the normalised coefficients.

The simulations were also run with two new sets of random phase-angles ($\beta_{n}^{m}$) in the stream function defined in Equation \ref{eq.stream}, and it was found that this does not produce a significant qualitative change in the results. Plots corresponding to figures \ref{fig.dir_r_dist} - \ref{fig.diff_coeffs_direction_norm} for the two additional realizations display the same trends as the figures shown in this paper. Actual values of the diffusion coefficients are broadly consistent. However in one of the additional realizations the ratio $\kappa_{\theta}^{\prime}/\kappa_{\phi}^{\prime} $ at the final time was equal to $1.6$ (while in the other additional realization this value was $0.1$). This appears to be related to a large deviation of the magnetic field lines, from the Parker spiral, in the $\theta^{\prime}$ direction. 

If the value of $N$ is increased from $N=50$ to $N=75$ it is found that the particles' trajectories are very similar to those in the $N=50$ case, with some small differences due to the fact that there is less power in the low $N$ modes.



\section{Discussion and Conclusions}
\label{sec.discuss}

The effect of large-scale IMF fluctuations on the propagation of SEPs is investigated by means of a full-orbit test-particle method. In the presence of large-scale turbulence, energetic particle populations experience more extensive cross-field transport than would be the case in a Parker spiral configuration, due to field line meandering \citep{1999AdSpR..23..581G}. 

In this paper, three propagation conditions were analysed and the cross-field transport in large scale turbulence characterized.
In scatter-free propagation SEPs simply follow field lines and the perpendicular transport observed is the result of magnetic field characteristics. 
By comparing populations at the same average radial distance from the Sun,
it was found that when pitch-angle scattering is present, the perpendicular transport is enhanced (see also Figure \ref{fig.xy_xz_stack}). Qualitatively, the introduction of scattering produces outlying particles that travel much further across the magnetic field than unscattered particles. The total number of outliers is small in these simulations, due to computing time constraints on the total number of particles considered, however they are representative of a population that can be found at large separation from the injection region. Observations show that the SEP fluxes measured at large angular separation from the source can be orders of magnitude smaller than those detected by a well connected spacecraft.

The introduction of scattering makes it possible for particles to jump onto nearby field lines, by a distance of the order of the Larmor radius, and, due to field line meandering, a superposition of a large number of these events can result in large cross-field displacement from the original field line.

The Parker spiral geometry poses some challenges to the interpretation of the simulations. In an attempt to resolve these challenges a local coordinate system was introduced to yield parallel and perpendicular displacements from the particle's original Parker spiral field line. The coordinate transformation to this Parker spiral system (as detailed in Appendix \ref{app.transform}) is generic and applicable to other heliospheric problems. For different scattering conditions, particles have travelled different distances along the field, influencing the calculated cross-field transport due to expansion of the magnetic field. It is important to take this effect into account in quantifying cross-field diffusion.

The method used here has been firstly, to analyse $\Delta s_{\perp}$ distributions for different scattering conditions after a fixed time from injection. It was found that the population which undergoes least scattering has the broadest distribution of $\Delta s_{\perp}$ values: this is because the least scattered particles can reach locations far away from their injection point where expansion and field line meandering cause large $\Delta s_{\perp}$ values.

Secondly, displacements were analysed at different times for the three regimes, with times chosen so that the average radial distance from the Sun of the population is $1$ AU. It was found that the $\lambda$=0.3 AU population has the broadest $\Delta s_{\perp}$ distribution, however this may be due to their having a significant number of particles at large distances from the Sun.

A normalisation of perpendicular displacements was introduced in an effort to minimise the effect of field line expansion. 
The normalisation only takes into account the Parker spiral expansion, thus the
expansion of the large-scale turbulence may not be completely compensated.
If the distributions of $\Delta s_{\perp} / \Delta s_{||}$ are plotted, it is seen that the population with $\lambda=2.0$ AU has the most cross-field transport per unit parallel transport, while the unscattered case has the least amount of cross-field transport per unit parallel transport. This appears to fit well with the qualitative conclusion reached from Figure \ref{fig.xy_xz_stack} that more distant outliers are present when  $\lambda$=2 AU compared to $\lambda$=0.3 AU.
While it is clearly established that perpendicular transport is enhanced in the scattering cases compared to the scatter-free one, the analysis is not conclusive as to which of the two scattering regimes results in the most efficient perpendicular diffusion, due to the challenge presented by field line expansion.

The ratio of perpendicular to parallel diffusion coefficients varies from 0.01 to 0.08 for the conditions analysed. Diffusion across the field in the latitudinal direction appears slower than in the longitudinal direction. This is consistent with SEP observations at high heliolatitude from Ulysses \citep{2003GeoRL..30sULY9D}. Transport in longitude could be aided by the electric field drift which is directed along $\mathbf{e}_{\phi}$.

\acknowledgments
We acknowledge support from the UK Science and Technology Facilities Council via a PhD studentship and standard grant ST/H002944/1.  
This work has received funding from the European Union Seventh 
Framework Programme (FP7/2007-2013) under grant agreement n.~263252 (COMESEP). We acknowledge use of the University of Central Lancashire's High Performance Computing Facility.

\appendix

\section{Appendix Material}
\subsection{Transformation to a local Parker Spiral coordinate system}
\label{app.transform}

A new coordinate system with origin at a point $(x_t, y_t, z_t)$ in space and one axis coinciding with the Parker spiral magnetic field direction at the point is introduced. The $\mathbf{e}_{l}$ axis points outwards along the local direction of the Parker spiral magnetic field, a second axis is in the direction of $\mathbf{e}_{\theta'}$=$-$$\mathbf{e}_{\theta}$ with $\mathbf{e}_{\theta}$ the standard spherical coordinate system unit vector and a third axis $\mathbf{e}_{\phi'}$  completes the orthogonal right-handed system. The latter axis does not coincide with the standard spherical coordinate system unit vector $\mathbf{e}_{\phi}$.

The angle between the radial direction and the direction of the magnetic field at the point is indicated by $\beta$ and is given by:
\begin{equation}
\beta = \tan^{-1} \left( \frac{-B_{\phi}}{B_r} \right) =\tan^{-1}\left( \frac{r\,\Omega_0 \, \sin{\theta}}{V_w}\right)
\label{eq.beta}
\end{equation}     
where the terms $B_{r}$ and $B_{\phi}$ are the radial and longitudinal components of the Parker spiral magnetic field, and can be found by setting $V_{\theta}=V_{\phi}=0$ in Eqs.~(\ref{eq.br}) and (\ref{eq.bphi}). 

If a vector has components $(v_x, v_y, v_z)$ in the cartesian coordinate system introduced in Section 1, its components $(v_l, v_{\phi'}, v_{\theta'})$ in the new Parker spiral coordinate system are given by:
\begin{equation}
\left[ \begin{array}{c} v_l \\ v_{\phi'} \\ v_{\theta'} \end{array} \right] =
\left[ \begin {array}{ccc} \sin{\theta}  \cos{\phi} \cos{\beta} +\sin{\beta} \sin{\phi}  & \sin{\theta}  \sin{\phi} \cos{\beta} -\sin{\beta} \cos{\phi}  & \cos{\beta} \cos{\theta}  \\ \noalign{\medskip} -\cos{\beta} \sin{\phi} +\sin {\theta}  \sin{\beta} \cos{\phi} &\cos{\beta} \cos{\phi} +\sin{\theta}  \sin{\beta} \sin{\phi} &\cos{\theta}  \sin{\beta} \\ \noalign{\medskip} -\cos{\theta}  \cos{\phi} &- \cos{\theta}  \sin{\phi} &\sin {\theta}  \end {array} \right] 
\left[ \begin{array}{c} v_x \\ v_y \\ v_z \end{array} \right]
\end{equation} \label{eq.transform}

\subsection{Determining the target point} \label{app.targetradius}

\begin{figure*}
\epsscale{0.8}
\plotone{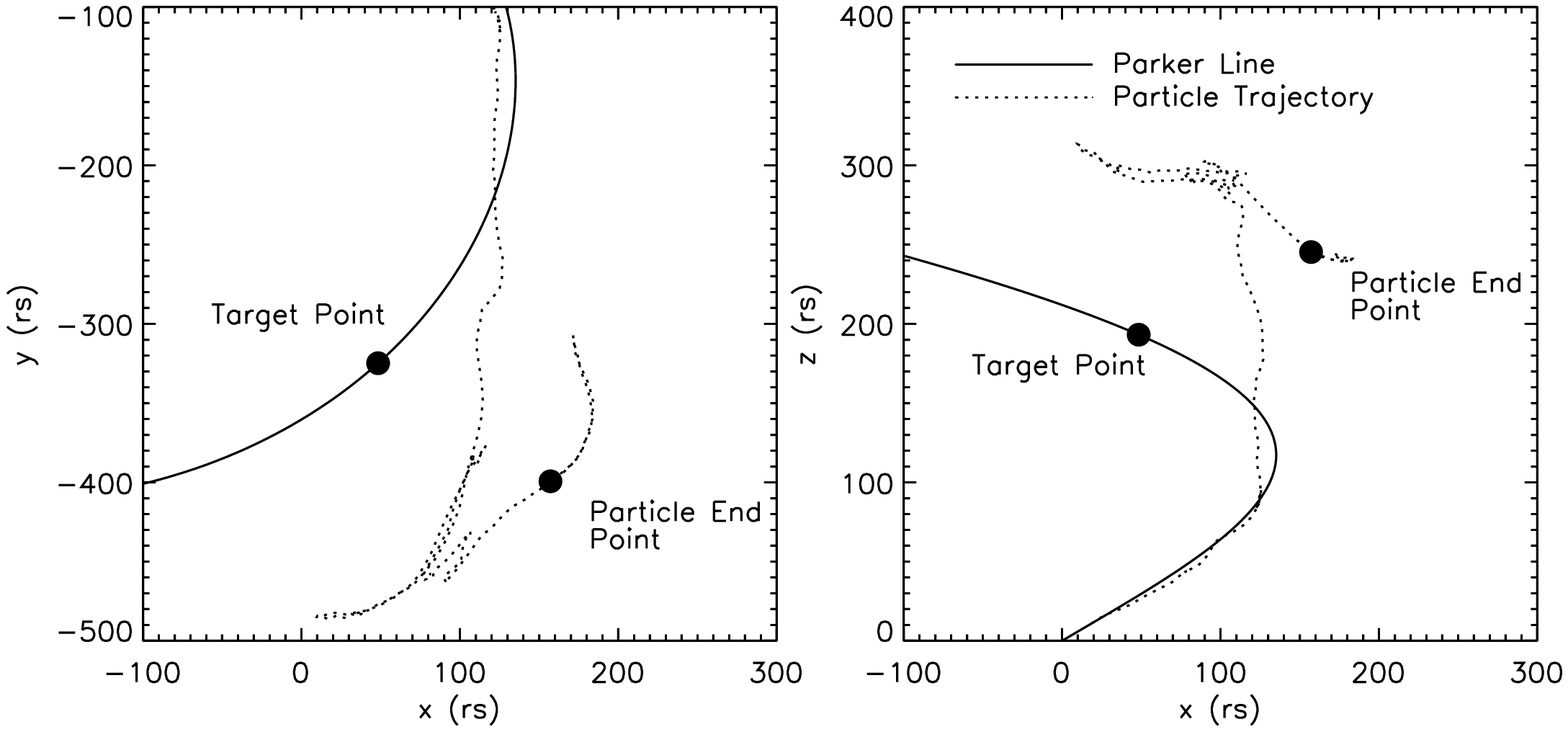}
\caption{An example particle trajectory (dotted) and the path of its original Parker spiral field line (solid) are shown, in the $x$-$y$ plane (left) and the $x$-$z$ plane (right). The particle's end point and target point are indicated by black dots. The target point was acquired by solving Equation \ref{eq.targ}, hence the perpendicular displacement of the particle from its original Parker line is simply the distance between its end point, and its target point.  \label{fig.target}} 
\end{figure*}

The target point is the location on the Parker spiral field line starting at the particle's initial position with the shortest distance to the particle's actual final position \citep{2011JGRA..11602102T} (see Figure \ref{fig.target}).

The target point is determined by means of an iterative procedure that differs from the one used by \citet{2011JGRA..11602102T}.
As an initial guess for the distance from the Sun of the target location, $r_t$, the value of the actual radial distance for the particle is used.
For a given $r_t$, 
the target longitude can then be found as $\phi(r_{t}) = \phi_{0} - (\Omega_0(r_{t}-r_{0}) / V_{w})$ given the particle's initial colatitude $\theta_{0}$, longitude $\phi_{0}$, and initial radius $r_{0}$. The target colatitude is  $\theta_{0}$.
If the cartesian components of this initial target location are indicated as 
$x_{t}$, $ y_{t}$ and $ z_{t}$,
the cartesian components of the $\Delta \mathbf{s}$ vector with respect to this target can be obtained.
Using the transform defined in Eq.~(\ref{eq.transform}), this can be converted into its components in the local Parker spiral coordinate system defined in Appendix \ref{app.transform}.

The optimal target location will have a zero component along the $\mathbf{e}_l$ axis and can therefore be found by solving the equation:
\begin{equation}
(\sin{\theta}  \cos{\phi} \cos{\beta} +\sin{\beta} \sin{\phi})[x_{e} - x_{t}] + ( \sin{\theta}  \sin{\phi} \cos{\beta} -\sin{\beta} \cos{\phi})[y_{e} - y_{t}] + (\cos{\beta} \cos{\theta})[z_{e} - z_{t}]=0
\label{eq.targ}
\end{equation}
where $(x_{e},y_{e},z_{e})$ is the particle's actual location. The equation is solved
by means of Brent's method \citep{Press:1993:NRF:563041}. This ensures that the particle's displacement with respect to the optimal target location is perpendicular to the Parker spiral.

Care must be taken when specifying the bracketing region to ensure that only one root of Equation \ref{eq.targ} lies within the region specified, and that this root corresponds to the closest target point to the particle end point.

\subsection{Normalization by scale length of Parker spiral flux tube}
\label{app.fluxnorm}

The cross-section of a Parker spiral flux tube increases with radial distance from the Sun. In this Section the scale lenghts characterising the cross-section in the $\phi$ and $\theta$ directions are determined, indicated as $L_{\phi}$ and $L_{\theta}$ respectively. These quantities are used to normalise perpendicular particle displacements so as to remove the effect of Parker spiral field line expansion.

The borders of the particle injection region, located at a distance $r_1$ from the Sun and with angular extent $\Delta \phi$ and $\Delta \theta$ in longitude and latitude respectively, define a Parker spiral flux tube with cross-section approximated by $\sigma_1$=$r_1^2 \, \Delta \phi \, \Delta \theta $. If the cross-section of the same flux tube at a distance $r$ from the Sun is indicated as $\sigma$, the conservation of magnetic flux requires that $\sigma$=$\sigma_1 B_1 / B$, where $B_1$ is the magnetic field magnitude at $r_1$ and $B$ its value at $r$.

In the Parker spiral:
\begin{equation}
B=\frac{B_{0}r_{0}^2}{r^{2}}\sqrt{1+\frac{r^2}{a^2}}
\end{equation}
where $a=v_{sw}/(\Omega_0\sin\theta)$. Hence:
\begin{equation}
\sigma=r^2 \, \Delta \phi \,\Delta \theta \, \sqrt{\frac{r_1^2+a^2}{r^2+a^2}}
\end{equation}
giving the following expressions for the cross-sectional scale lengths:
\begin{eqnarray}
L_{\phi} &=&  r \, \Delta \phi \, \sqrt{\frac{r_1^2+a^2}{r^2+a^2}} \\
L_{\theta}&=& r \, \Delta \theta
\end{eqnarray}

$L_{\phi}$ and $L_{\theta}$  are calculated using $r_1$=21.5 $r_s$ and $\Delta \phi$=$\Delta \theta$=$\pi/60$.



\bibliographystyle{apj}
\bibliography{biblio}

\end{document}